# New experimental evidence for Podkletnov effect

A.V. Fetisov[1]
fetisovav@mail.ru

## Abstract

In the works of E. Podkletnov et al. (for example [Physica C: Superconductivity, 1992, Vol. 203, p. 441]), a unique experiment is described in which a significant reduction in the weight of various test bodies was observed when placed above a rotating disk made of the superconductor $YBa_2Cu_3O_{6+\delta}$ (YBCO); the disk was in the superconducting state (at 20–70 K) and was subjected to an alternating (50 Hz–100 MHz) magnetic field. Numerous attempts to replicate this experiment in other laboratories have been unsuccessful; however, a broad discussion has been initiated regarding the possible nature of the "Podkletnov effect" – the weight loss of objects located above the superconductor. In the present experimental work, results similar to the Podkletnov effect have been obtained, which, however, occur at room temperature. During our experiments, gas-tight containers with the crystal hydrate $K_2CO_3{\cdot}1.5H_2O$ and dispersed YBCO exhibited a sharp loss of part of their weight during and after 0.5 hours of exposure to a magnetic field with a frequency of 50 MHz. Changes in the weight of various test bodies located in close proximity to the containers were recorded. It was found that these weight changes correlate with the external surface area of the bodies and do not depend on their chemical nature. A dependence of the magnitude of weight changes on the oxygen content in YBCO was obtained, which shows a significant dip in the region where superstructural orderings of the oxygen sublattice exist in the basal plane. This dip coincides in shape with the well-known dip in the dependence of $T_c$ on $\delta$ (the "60-K plateau"). The possible connection between these two extraordinary phenomena coexisting in YBCO compounds – inexplicable weight changes and high-temperature superconductivity – is discussed.

## 1 Introduction

Shortly after the discovery of cuprate high-temperature superconductors in 1986, articles began to appear describing various previously unobserved properties of these materials. Primarily, these were unusually high superconducting transition temperatures $T_c$ for compounds that had already been well-studied by then. However, in 1992, anomalous properties of a completely different sort for the high-temperature superconductor $YBa_2Cu_3O_{6+\delta}$ (YBCO) were reported in the paper of E. Podkletnov et al. [1]. The authors claimed that a $SiO_2$ sample lost up to 0.3% of its weight when placed above a YBCO disk rotating in a travelling magnetic field. An unstable regime was also observed, in which the sample weight became either greater or less. The data obtained also indicated a de-crease in the weight of the superconducting disk itself. It was noted that the rotation speed of the disk and the frequency of the magnetic field were important parameters affecting the weight loss. Later, additional information was provided about weight loss of up to 2.1%, which could be achieved for any samples made of metal, plastic, ceramics, etc. [2]. The weight changes observed in [1, 2] were subsequently referred to as the "Podkletnov effect". Meanwhile, numerous attempts by other authors to reproduce these results have been unsuccessful (see, for example, [3, 4]).

[1] Institute of Metallurgy of the Ural Branch of the Russian Academy of Sciences, Ekaterinburg, Russian Federation

There have been attempts at theoretical justification for the weight change of objects located above a rotating superconductor in a magnetic field (see, for example, [5, 6]). Some attempts were based on the description of the gravitational shielding effect in the frame of the quantum theory of General Relativity due to the presence of a superfluid [5], others – on the hypothesis that magnetic vortices within type-II superconductors may interact with the gravitational field, introducing distortions into it [6]. All attempts made so far have been strictly related to the superconducting state of the material, and the case of the normal state has not yet been considered. On the other hand, precision measurements of the weight of various superconductors have not shown any change in this parameter up to $\sim 2 \cdot 10^{-4}$ wt.% during the transition through $T_c$ [7]. Other precision measurements conducted at low temperatures [8] have not revealed any emergence of energy fields near a rotating superconducting disk. These data may suggest that neither the superconducting state itself nor rotation in this state are determining factors for the Podkletnov effect.

Previously, we discovered that some YBCO samples change in weight upon hydration by an amount not commensurate with the amount of absorbed water [9–12]. In other words, part of this weight change was not related to the hydration process; its nature has not been established. The low-temperature magnetic properties were also intriguing. Before hydration, this material is a superconductor with a $T_c$ characteristic for this chemical composition. However, upon hydration, a Meissner paramagnetic effect begins to manifest itself and does not disappear even in strong magnetic fields. At temperatures $T > T_c$, up to 300 K, YBCO samples hydrated to about 2 wt.% exhibit diamagnetism with $M = -1.34 \cdot 10^{-3}$ Oe (at $T = 300$ K; $H = 10^4$ Oe). A distinctive feature of the studied samples was that during their preparation they were carefully protected from atmospheric moisture except for a brief final period when they were exposed to air with humidity $p_{H_2O}$ = 110±10 Pa. In this work, we continue to study changes in weight of YBCO during its hydration. It turned out that the magnitude of such changes is determined by external electromagnetic fields. After exposure to these fields, the weight of the samples begins to change. Our results are very similar to those described by E. Podkletnov and could clarify and supplement them with new information.

## 2 Experimental Details

The YBCO material with tetragonal crystal symmetry was prepared using the ceramic technology described in [9–12]. The saturation of the YBCO lattice with oxygen was carried out in dried air or in an oxygen atmosphere (produced using an Invacare Perfecto$_2$ oxygen concentrator) for 5 hours at temperatures ranging from 525 to 890 °C, followed by quenching on a massive ceramic plate. The sintered material was ground in an agate mortar and placed into quartz crucibles with an inner diameter of 7.3 mm and a height of 13 mm, with 750±50 mg of powder in each crucible. All procedures were conducted in a box with a dried atmosphere. The crucibles containing the YBCO material were exposed to air for some time at $p_{H_2O}$ = 110±10 Pa, after which they were placed into gas-tight glass containers with the crystal hydrate $K_2CO_3 \cdot 1.5H_2O$ – two crucibles per container (hereafter referred to as "reactor"). The presence of the crystal hydrate in the reactors maintained a constant relative humidity of 43%. The volume of each reactor was 15 ml, and its cap was made of low-density polyethylene. Reactors containing YBCO samples will be referred to as reactors S. An identical reactor without samples, which was used as a "reactor for comparison", will be referred to as R; it underwent all stages of weighing and intermediate storage alongside reactors S (while being kept at a sufficient distance from them to avoid any influence). The initial weight of the reactors was determined approximately 5 minutes after placing the samples inside, using analytical scales Shimadzu AUW-120D with an accuracy of 0.02 mg. Each weighing procedure consisted of eight separate weighings conducted with reactors R and S in the sequence: {R, S} ×4 times. The current weight changes of reactor S were determined relative to the weight of reactor R ($W_R$): $\Delta W_S = (\Sigma W_S - \Sigma W_R)/4 - (\Sigma W_S^{initial} - \Sigma W_R^{initial})/4$. Before each weighing procedure, the scales were calibrated using internal calibration weights. After the initial weighing, each reactor S was placed inside a solenoid, where it was subjected to an alternating magnetic field at a frequency of 50 MHz for 30 minutes. The measured radiation power at the exposure site for the reactors was approximately 4 μW/sm². The subsequent two weighing procedures for the reactors were conducted 45 and 90 min. after the initial weighing. Thereafter, this procedure began to be measured daily. In between weighings, the reactors were stored in steel safes at a temperature of 25.0±0.5 °C – each reactor was kept in a separate safe in a tin-plate box and was additionally shielded from thermal and electromagnetic radiation by the walls of an aluminum box.

X-ray structural analysis (XRD) was performed on a Shimadzu XRD-7000 diffractometer (CuKα radiation, Bragg angle range 2Θ = 10 ÷ 80°, step size 0.02°, and exposure time at a point 2 s). The analysis of the crystal structure was carried out using the GSAS software package [13], based on the crystal structure model presented in [14].

## 3 Experimental Results and Discussion

### *3.1 Material characterization by X-ray diffraction*

The results of X-ray diffraction (XRD) obtained on YBCO samples subjected to various treatments during oxidative annealing are presented in Figs. 1 and 2, and Table 1. It should be emphasized that the samples quenched from the annealing temperature exhibit the standard dependence of lattice parameters on the oxygen index δ: as δ decreases, a smooth reduction in the orthorhombic distortion of the YBCO lattice is observed until a transition to the tetragonal phase occurs at δ ≈ 0.35. This can be clearly seen in Fig. 2 by the shift of characteristic peaks in the X-ray diffraction patterns in the range 2θ = 44 – 50 degrees.

To determine the oxygen content in the samples, we utilized the well-established fact regarding its correlation with the lattice parameter *c* – its dependence on *c* does not significantly deviate from linear (see, for example, [14–16]). To obtain the required dependence, we used data on δ and *c* from the work [14]:

$$\delta = 5.7837 \cdot (11.834 - c). \quad (1)$$

As justification for our chosen method of determining δ, we compare the value of δ obtained from equation (1) for a sample annealed at 585 °C in air (δ = 0.726, see Table 1) with the value previously found by us [9] using iodometric titration for a sample prepared by the identical method (δ = 0.75 ± 0.02). It can be stated that the obtained values are quite close.

**Table 1** XRD characterization of YBCO samples subjected to various treatments during oxidative annealing

| | Annealing temperature / atmospheric condition | | | | | | |
|---|---|---|---|---|---|---|---|
| **Lattice parameters** | 530 $^0$C / $O_2$ | 525 $^0$C / air | 585 $^0$C / air | 650 $^0$C / air | 725 $^0$C / air | 815 $^0$C / air | 890 $^0$C / air |
| *a*, Å | 3.82153 | 3.82042 | 3.82699 | 3.82846 | 3.83510 | 3.85045 | 3.86019 |
| *b*, Å | 3.88681 | 3.88339 | 3.88223 | 3.88029 | 3.87550 | 3.86042 | 3.86019 |
| *c*, Å | 11.68681 | 11.69307 | 11.70856 | 11.72075 | 11.73292 | 11.75370 | 11.76958 |
| *V*, Å$^3$ | 173.591 | 173.481 | 173.957 | 174.118 | 174.386 | 174.711 | 175.379 |
| $R_{wp}$*, % | 8.010 | 9.554 | 9.725 | 8.671 | 7.849 | 8.863 | 9.586 |
| δ according to [14] | 0.851 | 0.815 | 0.726 | 0.655 | 0.585 | 0.464 | 0.373 |

* $R_{wp}$ is the weighted profile R-factor

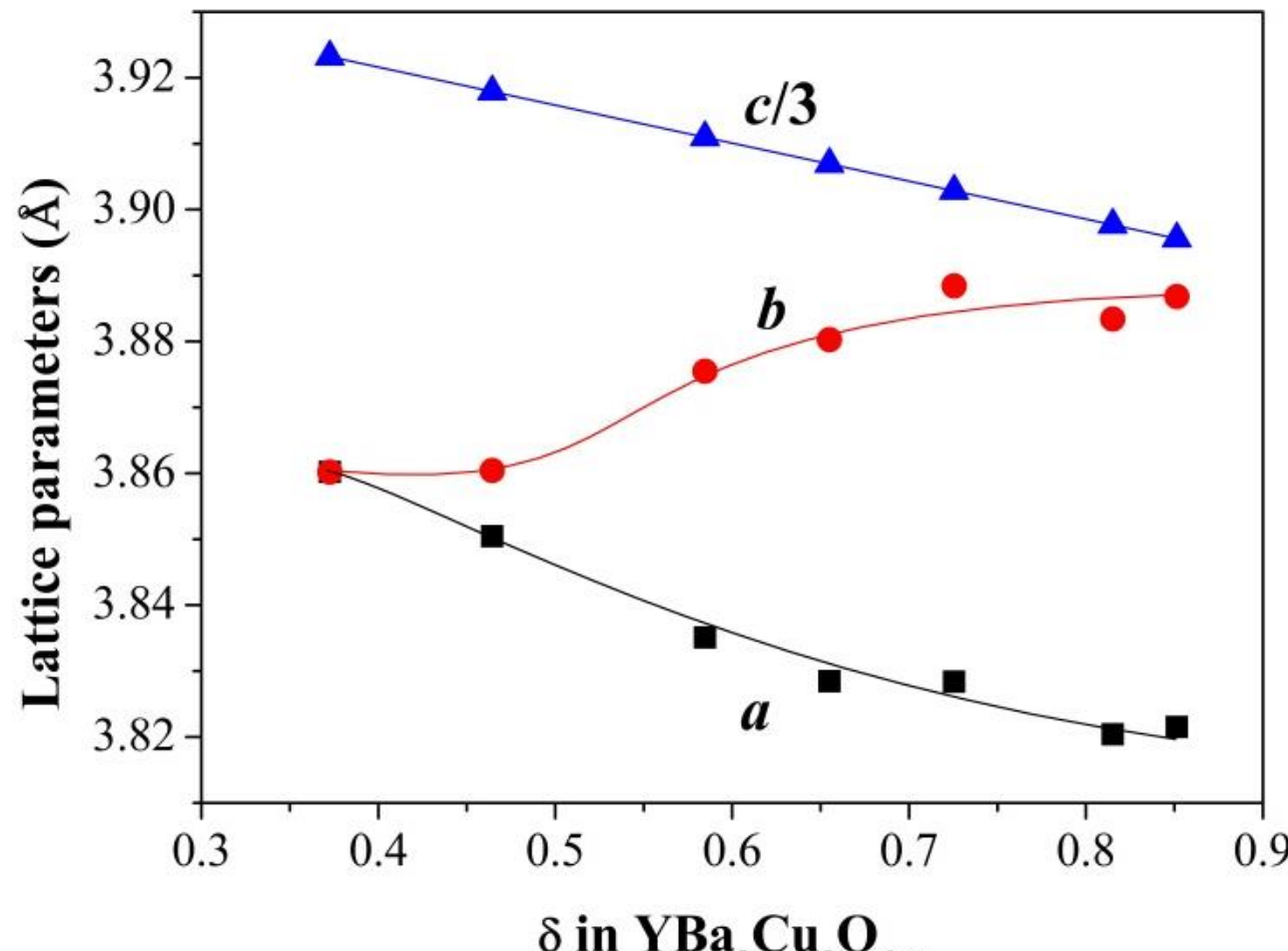

**Fig. 1** Graphical representation of the dependence of lattice parameters on oxygen content in YBCO.

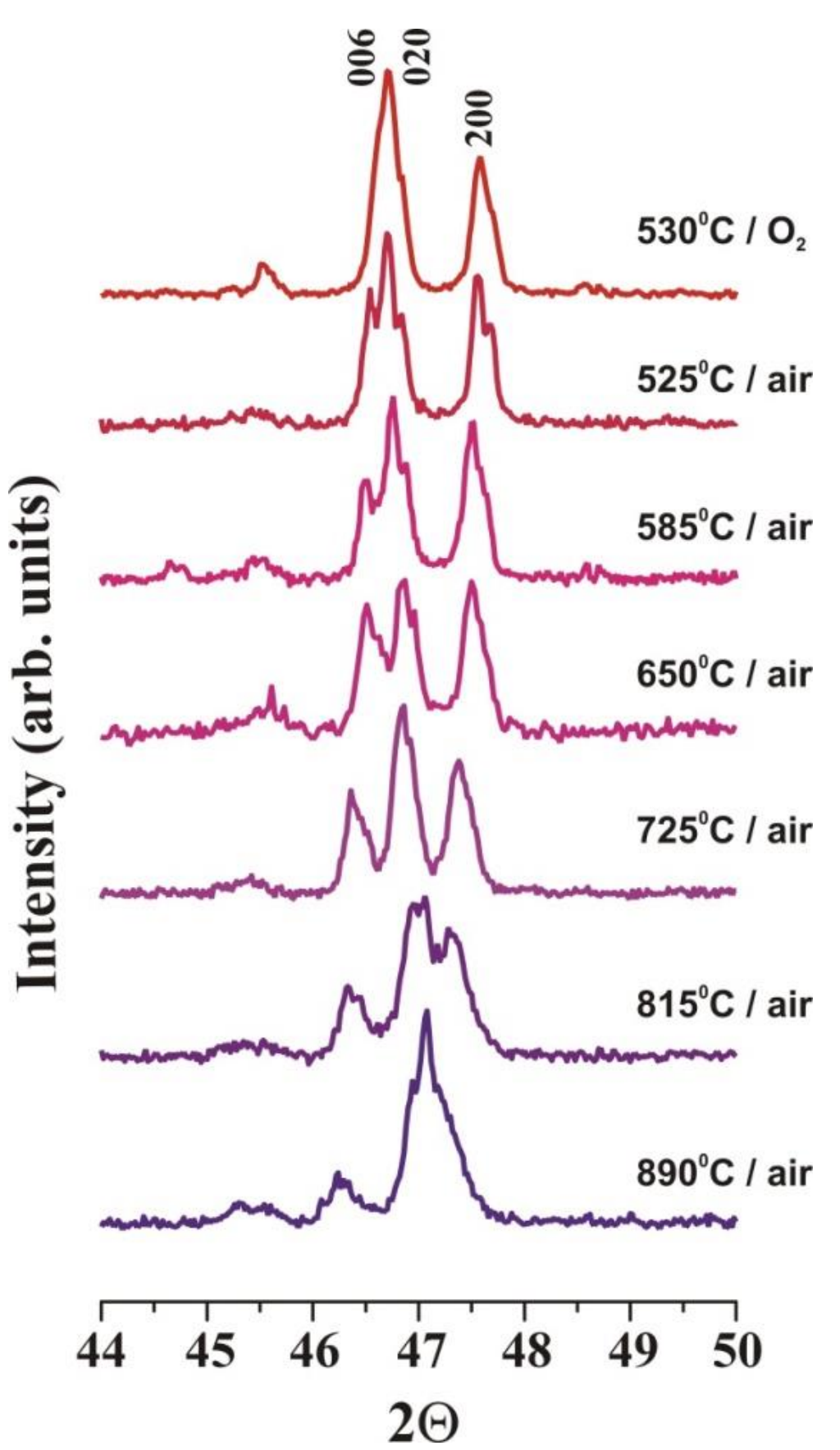

**Fig. 2** The profiles of the triplet 200, 020, and 006 taken at room temperature of YBCO samples with different oxygen treatment.

### *3.2 Instability in the weight of reactors with YBCO samples*

According to the Podkletnov's work [2], the optimal frequency of the processing field at which the weight loss effect reaches its maximum is 3.5±0.3 MHz. In our preliminary experiments it was found that the maximum effect is achieved at frequencies of 50 MHz and above. So, they were the specific frequencies of the alternating magnetic field that we used in our study for the treatment of the reactors (see paragraph 2).

The results of weight measurements on reactors with various samples and under different experimental conditions are presented in Fig. 3. It can be seen that during and after field treatment, the reactors containing YBCO superconductor experience a significant weight loss. The reduction in weight is primarily observed during the first 1.5 hours of the experiment. Additionally, Fig. 3 presents results of a 4-min. treatment of a reactor from a magnetron at a frequency of 2 GHz conducted from a distance of 1 m (curve *5*). As one can see, the weight loss for the reactor subjected to that irradiation is at the same level as the weight loss observed on the frequency 50 MHz. On the other hand, without treatment of the reactor with a field, the resulting weight loss becomes less significant (curves *3* and *3'*).

Fig. 3 also shows dependencies (curves *1* and *2*) demonstrating that the observed weight loss is a consequence of the interaction between YBCO samples and water vapor emanating from $K_2CO_3 \cdot 1.5H_2O$ crystal hydrate. Meanwhile, this reaction is "contained" within closed volume, subsequently, the observed changes in weight of reactors cannot be explained solely on this basis. To elucidate the nature of these mysterious weight changes is the main issue to be solved in this study and maybe in our future research.

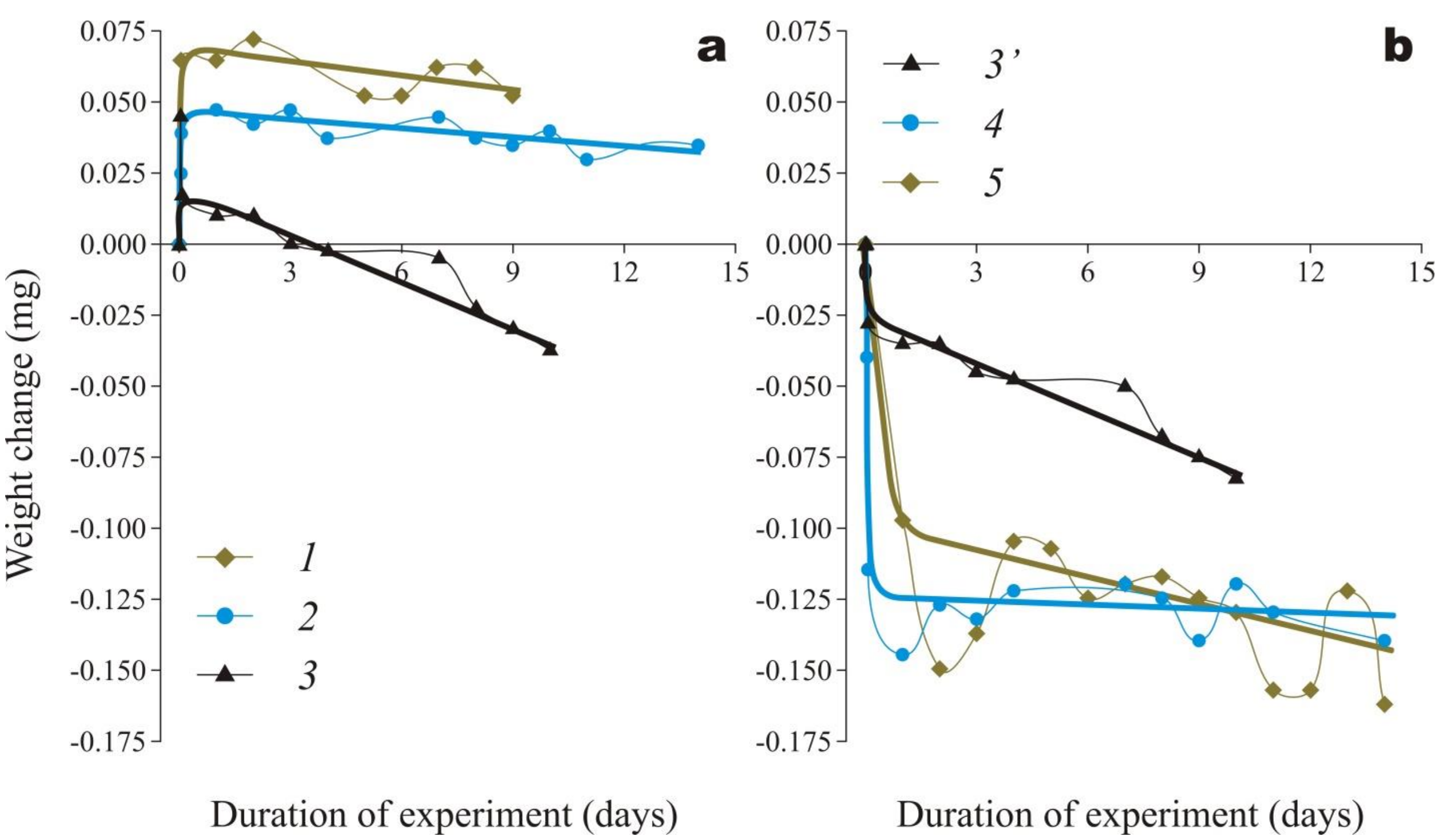

**Fig. 3** Changes in weight of reactors after loading various samples into them. Oxidative annealing of YBCO samples was carried out at 525°C in air;
**a**: **curve *1*** corresponds to the "standard experiment" carried out in our study − with YBCO samples and field treatment at a frequency of 50 MHz, − but instead of crystal hydrate, dehydrated silica gel was in the reactor; **curve *2*** corresponds to the standard experiment, but instead of YBCO samples, $SiO_2$ samples were used; **curve *3*** corresponds to the standard experiment without treatment by electromagnetic field;
**b**: In this figure, all curves are plotted by subtracting the contribution of the initial jump $\Delta W = +0.045$ mg revealed in the experiment with the $SiO_2$ sample, which is a consequence of bringing the reactors to ambient conditions; **curve *3'*** corresponds to the standard experiment without treatment by electromagnetic field; **curve *4*** corresponds to the standard experiment with field treatment at 50 MHz; **curve *5*** corresponds to the standard experiment with field treatment at 2 GHz.

Figure 4 shows kinetic curves of the weight loss ($\Delta W_S$) obtained on the reactors with YBCO samples having different oxygen content. It is easy to see that in the region of intermediate $\delta$ values, the magnitude of the reactor weight loss is significantly less than in the case of low and high $\delta$ values. The resulting "dip" is more clearly seen in Figure 5, where the $\Delta W_S$ data are accompanied by information on the structural features of YBCO depending on its oxygen composition.

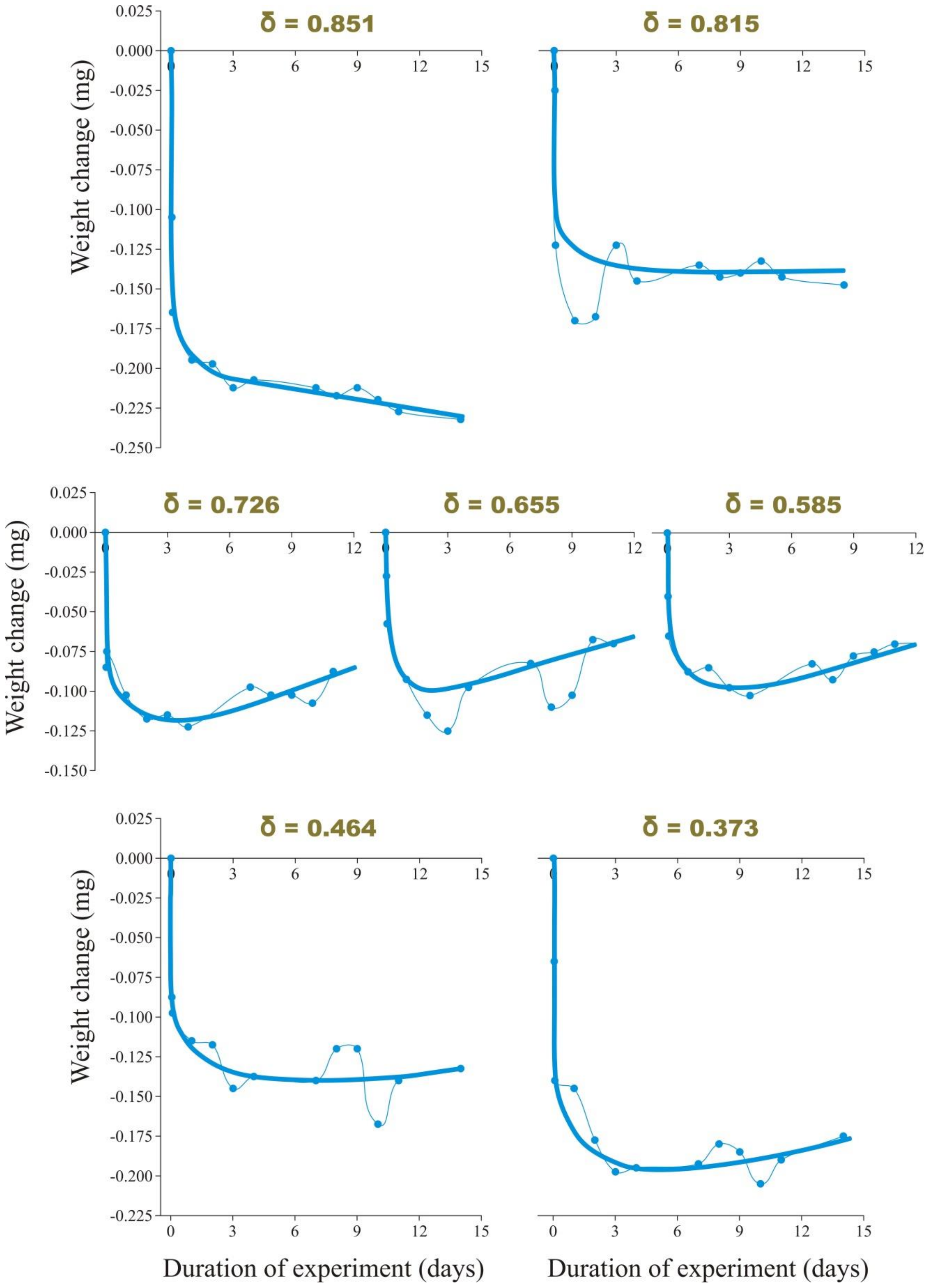

**Fig. 4** The effect of weight loss of reactors with $YBa_2Cu_3O_{6+\delta}$ samples in depending on the oxygen index $\delta$.

Comparison of the data on the structure of YBCO with the data on the "drop effect" of $\Delta W_S$ shows that the decrease in $\Delta W_S$ observed on the low-oxygen side coincides with the increase in orthorhombic distortion of YBCO after its tetragonal-to-orthorhombic (T→O) transition at $\delta \approx 0.35$ (see also Fig. 1). According to [17, 18], a structural phase of the ortho-II type with a specific superstructural oxygen ordering in the basal plane is formed in this region, whose existence range extends up to $\delta \approx 0.65$. For values $0.60 < \delta < 0.82$, there is also a region where other superstructural orderings in the oxygen sublattice are observed, namely, the ortho-V, ortho-VIII, and ortho-III phases. The sharp increase in the drop effect on the high-oxygen side at $\delta > 0.82$ coincides with the formation region of the ortho-I phase. Thus, it can be stated that the magnitude of the drop effect is apparently controlled by the structural features of YBCO; minimal values of $|\Delta W|$ are observed for the superstructural phases ortho-V and ortho-VIII.

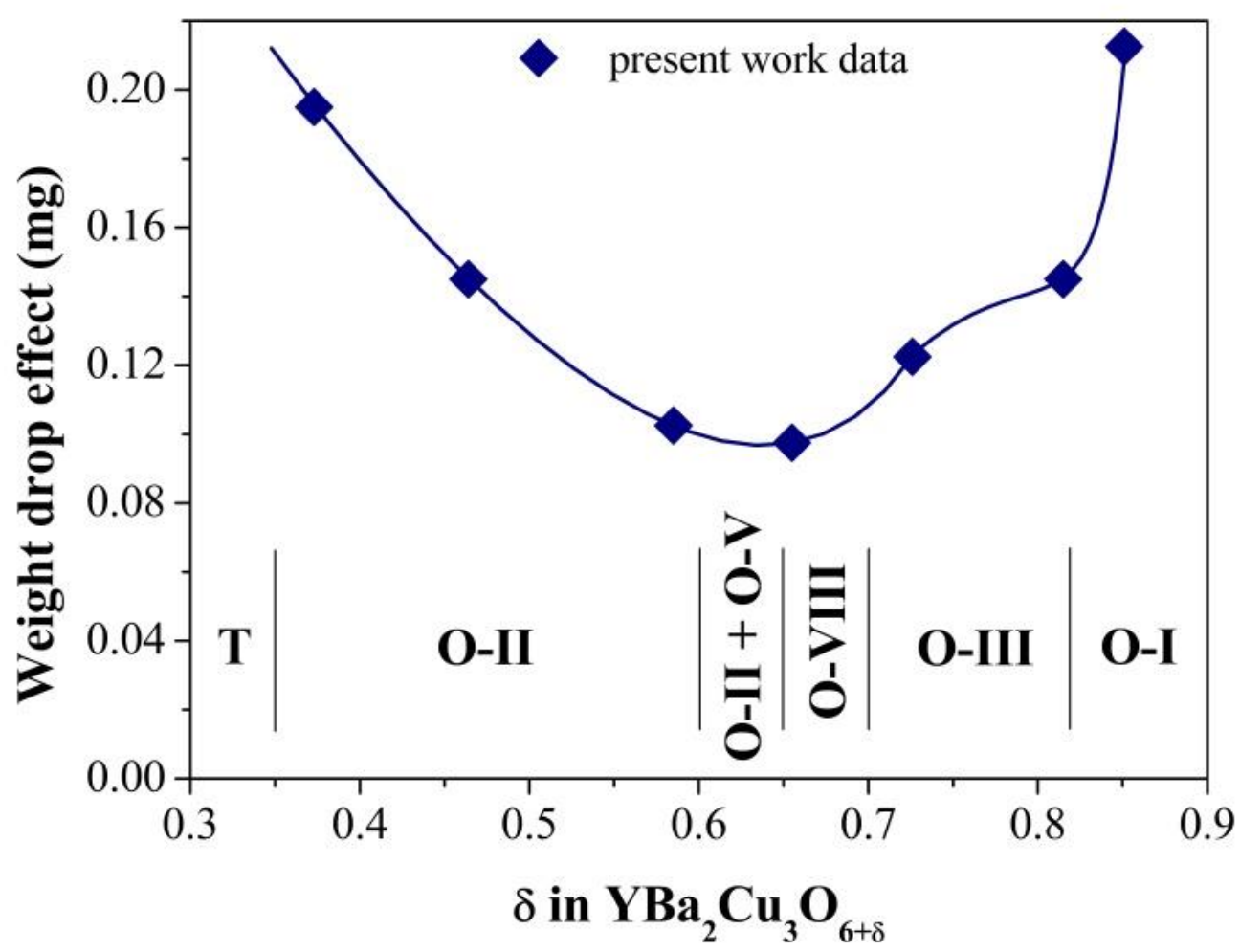

**Fig. 5** The weight loss of reactors with $YBa_2Cu_3O_{6+\delta}$ samples ($\Delta W_S$) in depending on oxygen index $\delta$. The correspondence of the characteristic areas $\Delta W_S$ to the structural features of YBCO according to the data [17].

Interestingly, the same correspondence to the structural features of YBCO is observed for its transition temperature of superconducting state, $T_c$. The dependence $T_c(\delta)$ of this superconductor exhibits a characteristic "60-K plateau", which, as noted, begins with the emergence of the O-II superstructure and ends with the formation of the O-I phase [19, 20]. This dependence, taken from several studies, is depicted in Fig. 6. It is not difficult to notice the similarity in shape and position of the 60-K plateau and the bottom of the $\Delta W(\delta)$ dependence.

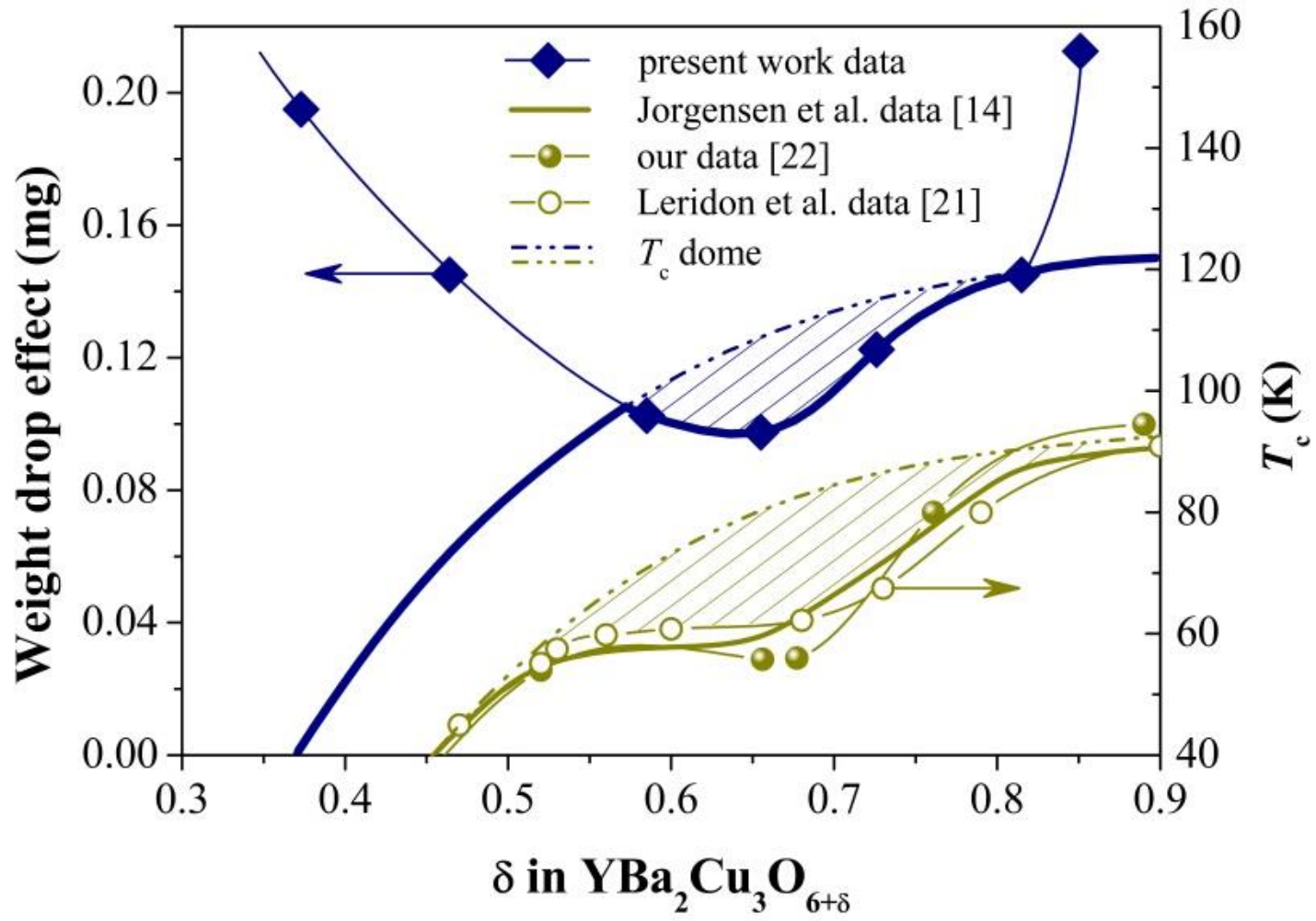

**Fig. 6** On the possible reason for the "dip" in the $T_c(\delta)$ dependence for YBCO.

It has been shown [23, 24] that the dependence $T_c(\delta)$ corresponding to YBCO aligns with the dome-shaped dependence $T_c(p)$ characteristic of all cuprate high-temperature superconductors, where $p$ is the concentration of free charge carriers responsible for superconducting current. Thus, the "60-K plateau" is actually a "dip" in the $T_c$-dome[2]. In this work, we propose to consider the hypothesis that the dip in the $T_c(\delta)$ dependence results from a dip in the $\Delta W(\delta)$ dependence. This leads to an interesting consequence:

Typically, the dip (or 60-K plateau) in the $T_c(\delta)$ dependence of YBCO superconductors is explained by the existence of some unfavorable factor manifesting in its electron-hole subsystem within a corresponding $\delta$ range. This could be either a decrease in concentration $p$ due to occurring superstructural orderings [19, 25-28] or the influence of the "1/8 effect" [15, 29] when reaching $p = 1/8$. There are also notions about certain "magic numbers" for $p$ [30], as well as about overdoping of apical oxygen at $\delta = 0.5–0.6$ [31], which according to authors should lead to a decrease in $T_c$ and consequently to a plateau in the $T_c(\delta)$ dependence. In these works, the unfavorable factor influences the existing mechanism of high-temperature superconductivity (HTSC) and, as a result, the $T_c$-dome. Now let us consider what would happen if we assume that there is no unfavorable factor and that the original dependence $T_c(\delta)$ is formed by two competing factors: the $T_c$-dome and some function $T_c = f(\Delta W(\delta))$, which resembles $\Delta W(\delta)$. In Fig. 6, it can be seen that they compete for a minimum in values of $T_c$. Such a situation is possible if processes determining the shape of both the dome and $\Delta W(\delta)$ are links in one more general process and parameter $\delta$ (or $p$) affects their intensity differently (an analogy for this situation could be oxidation of a solid by oxygen of a gas phase when at some temperature process rate is controlled by interfacial reaction while at another it is controlled by diffusion of oxygen within solid). Given that the processes determining the shape of the $T_c$-dome are fundamental in mechanism of high-temperature superconductivity[3], it follows that the processes responsible for the weight drop effect are also participants in HTSC mechanism. It should be noted that for other HTSC compounds where significant oxygen non-stoichiometry is absent, function $T_c = f(\Delta W)$ should not depend on composition, even in cation sublattice (which we will verify in future work). This may explain why these compounds exhibit a $T_c$ dome without a dip.

### *3.3 Instability in the weight of test bodies placed near reactor with YBCO powder*

Figures 7–9 and Table 2 present the scheme and results of the influence of reactors containing YBCO samples on the weight of various test bodies located in close proximity to them. In this experimental setup, both the reactor placed in a tin-plate box and the test body were positioned on high stands made of thin cardboard to maintain a significant distance from other objects. The center of the test body was located below and to the side

---

[2] Here it is taken into account that the charge concentration $p$ is a function of $\delta$, close to linear [15, 22].
[3] It should be noted that the process itself is currently not fully understood.

from the reactor at a distance of 12–13 sm from its center. When determining the weight of the bodies, changes in buoyancy due to variations in temperature and air pressure were taken into account:

$$\Delta W^{P,T} = V \cdot \left(0.003974 \cdot \Delta T - 1.17 \cdot {}^{\Delta P}/_{P_0}\right), \qquad (2)$$

where $V$ is the external volume of the reactor [ml]; $\Delta T$ [°C] and $\Delta P$ [kPa] are the changes in temperature and pressure that occurred from the initial moment in time when $T = T_0$ and $P = P_0$; the coefficients used in the equation express the magnitude of buoyant force for unit values of $V$, $\Delta T$, and $\Delta P$.

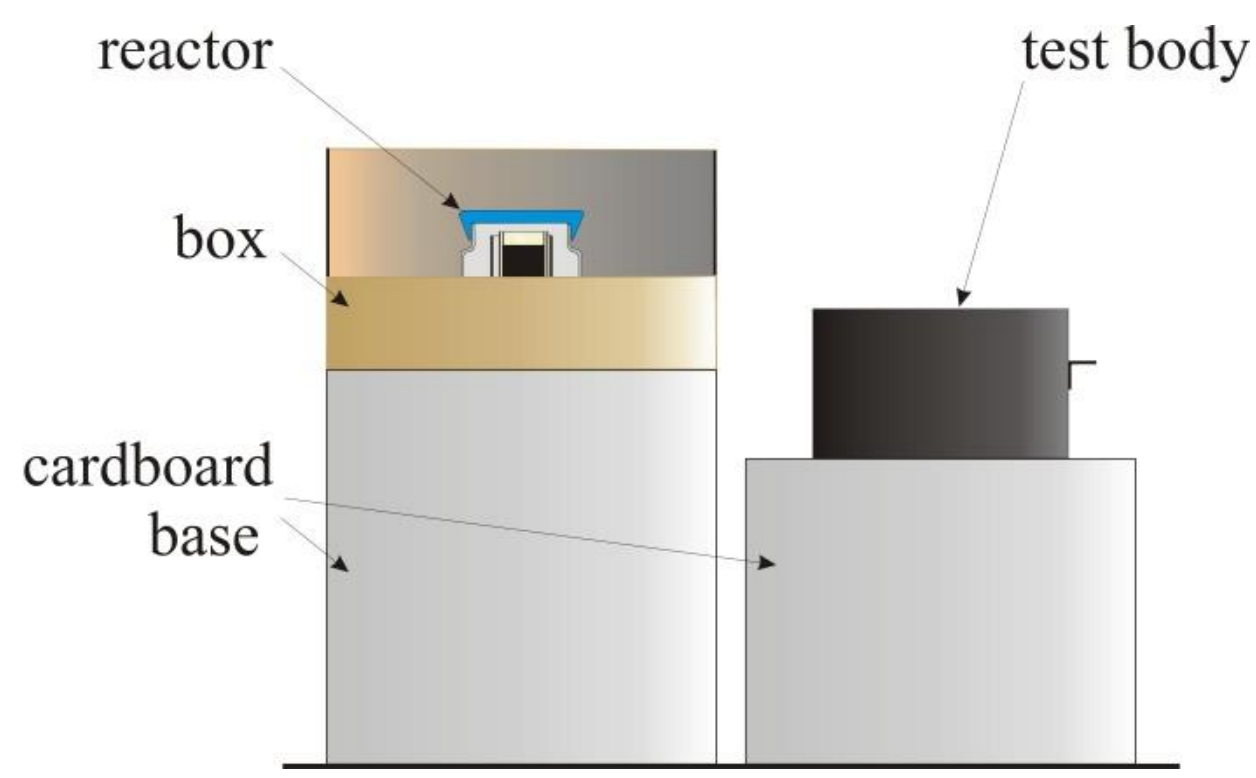

**Fig. 7** The position of the test body relative to the reactor with samples in the experiments presented in Fig. 8.

**Table 2** Characterization of the test bodies involved in the experiments presented in Fig. 8

| Material of test body /Figure | Weight, g | Area of surface, sm$^2$ | Maximum effect of weight change, mg |
|---|---|---|---|
| Tin-plate 1/Fig. 8a | 21.3 | 258 | 0.177 |
| Aluminum/Fig. 8a | 39.5 | 195 | 0.118 |
| Copper/Fig. 8b | 21.8 | 16 | 0.044 |
| Plastic/Fig. 8b | 10. 6 | 120 | 0.087 |
| Tin-plate_2/Fig. 8c | 28.5 | 312 | -0.241 |
| Tin-plate 3/Fig. 8d | 19.5 | 238 | -0.125 |

As can be seen in Fig. 8, the weight of each test body located near the "operating" reactor begins to change impulsively, regardless of its chemical composition. A regularity is observed: the larger the external surface area of the body, the greater the magnitude of the impulses, as shown in Fig. 9. Primarily, positive impulses are observed; however, when the surface area of the body exceeds 300 sm$^2$, the impulses become predominantly negative. At this point, the weight drop of the reactor itself noticeably increases (see Fig. 8c). A similar increase in ΔW of the reactor is also observed when a second reactor is placed next to it instead of a test body, as seen in Fig. 8c, curves *2* and *2'*.

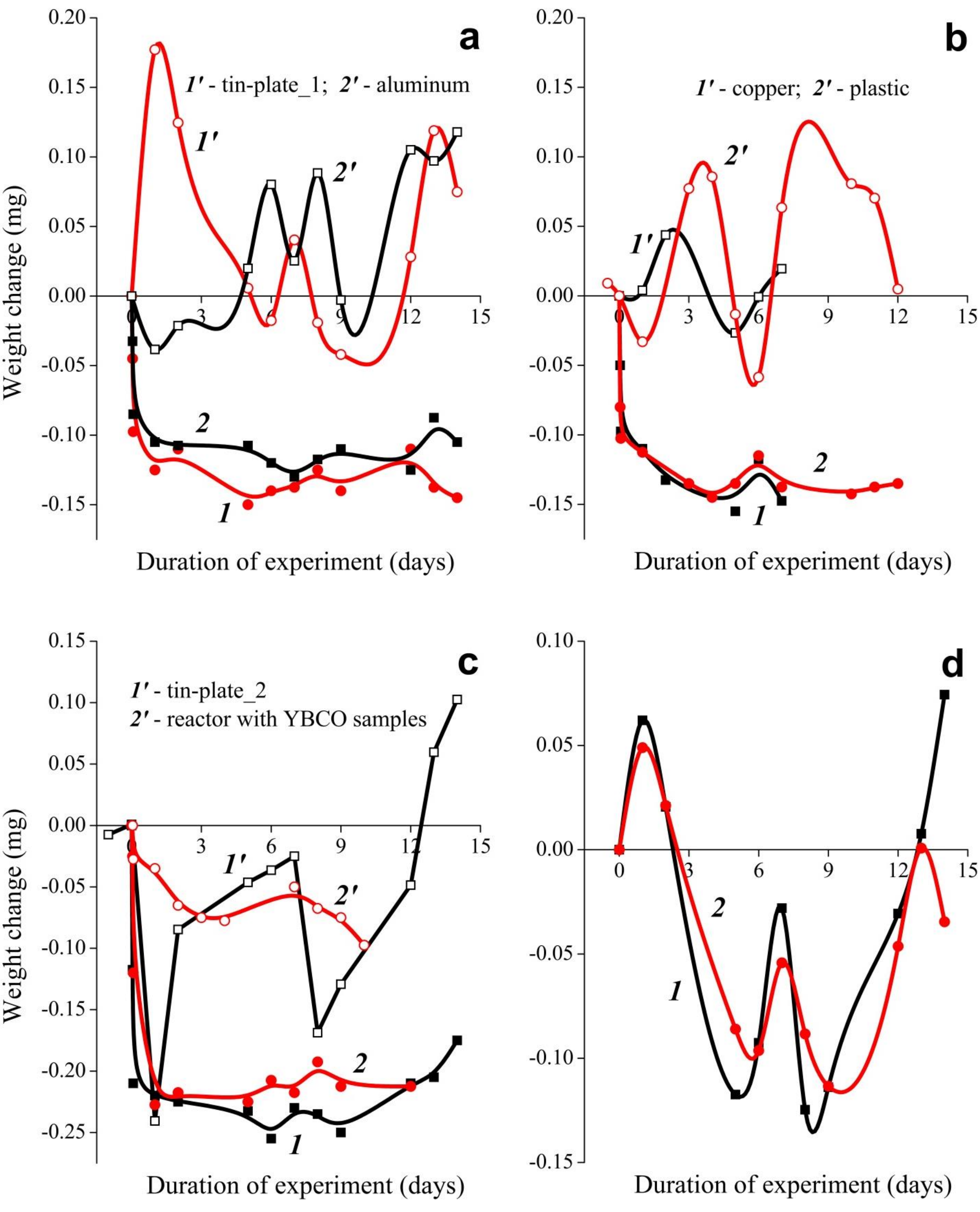

**Fig. 8** **a–c** – Changes in the weight of various bodies located in the immediate vicinity of the reactors with YBCO samples (curves *1′*, *2′*) compared to the change in the weight of the reactors themselves (curves *1*, *2*); **d** – changes in the weight of the box under the reactor with samples, made of tin-plate (curve *1*) compared to the sum (curve 2): $0.7\Delta W_1 + 0.6\Delta W_2$, where $W_1$ is the change in the weight of the reactor itself, see Fig. a, curve *1*, and $W_2$ is the change in the weight of the body made of tin-plate located nearby, see Fig. a, curve *1′*.

Changes in the weight of the box under one of the reactors were also recorded, as shown in Fig. 8d, curve *1*. It turned out that these changes are a superposition of the weight changes of the reactor and the nearby body. Therefore, it can be suggested that one part of the box (presumably located closer to the reactor) acts as a "weight donor", like the reactor, while another part serves as a "weight acceptor", like the test body.

**Fig. 9** The influence of the test body surface area on the absolute value of the change in its weight during its stay near the reactor with YBCO powder.

In the works of E. Podkletnov, various bodies located above the YBCO superconductor lost a portion of their weight. When the bodies were positioned to the side, outside the vertical projection of YBCO, this effect was absent. In our experiments on the influence of YBCO powder on nearby bodies, this condition was not significant. This demonstrates that, at least in our case, the weight drop effect is not related to a modification of gravity.

Summarizing all of the above, it becomes evident that existing knowledge about YBCO, particularly regarding its structural plane CuO, is incomplete. At present, however, it is difficult to propose any suitable model that would realistically describe the experimental results obtained in our work. Nevertheless, let us hypothesize that there exists an as-yet unstudied matter (M) within this plane that affects weight and possesses some unique properties. During our experiments, it is displaced by water first into the reactor and then outward through the plastic cap. Outside the reactor, M tends to adsorb onto the surface of nearby bodies, displacing any adsorbed water in the process. The bodies may either gain or lose weight depending on the mass ratio of M to desorbed water (the mechanism of weight change may be more complex) [4].

Regarding the similarities between $T_c(\delta)$ and $\Delta W_S(\delta)$ dependencies, empty –Cu–$V_O$– structural chains present in the superstructural phases of YBCO may play a certain role here. A significant portion of the M-phase may be blocked by high walls of neighboring –Cu–O– chains from participating in both the process of displacing M by water (in the case of the $\Delta W_S$-effect) and in high-temperature superconductivity processes.

---

[4] In Podkletnov's experiments, the process of expelling M from YBCO apparently occurred under the influence of the centrifugal force of a rotating disk. After that, under the influence of a magnetic field, M moved strictly upward.

## 4 Conclusion

For the dispersed material $YBa_2Cu_3O_{6+\delta}$ (YBCO) enclosed in gas-tight containers with $K_2CO_3 \cdot 1.5H_2O$, a sharp and significant decrease in weight was observed after exposure to an alternating magnetic field with a frequency of 50 MHz. A key feature that enabled this result was the maintenance of a non-humid atmosphere at all stages of YBCO preparation, with a final exposure to an atmosphere with $p_{H_2O}$ = 110±10 Pa during the last stage. A dependence of the weight loss on the oxygen content δ in YBCO was obtained, which exhibits a strong dip at intermediate values of δ, resembling the well-known dip in the $T_c$ vs. δ dependence ("60-K plateau"). Changes in weight were also recorded for various bodies located in close proximity to the containers with YBCO material. A hypothesis has been proposed regarding the connection between the difficult-to-explain weight changes in YBCO and its high superconducting transition temperature.

The author expresses gratitude to colleagues O.M. Fedorova and T.I. Filinkova for providing X-ray diffraction data of YBCO samples.

## References

1. Podkletnov E., Nieminen R. A possibility of gravitational force shielding by bulk $YBa_2Cu_3O_{7-x}$ superconductor. *Phys. C: Supercond.* **203** (1992) 441–444, https://doi.org/10.1016/0921-4534(92)90055-H
2. Podkletnov E. Weak Gravitational Shielding Properties of Composite Bulk $YBa_2Cu_3O_{7-x}$ Superconductor Below 70 K under an EM Field, MSU-chem-95-cond-mat/9701074 5 Feb 1997.
3. Unnikrishnan C.S. Does a superconductor shield gravity? *Phys. C.: Supercond.* **266** (1996) 133–137, https://doi.org/10.1016/0921-4534(96)00340-1
4. Hathaway G., Cleveland B., Bao Y. Gravity modification experiment using a rotating superconducting disk and radio frequency fields. *Phys. C.: Supercond.* **385** (2003) 488–500, https://doi.org/10.1016/S0921-4534(02)02284-0
5. Modanese G. Role of a "local" cosmological constant in Euclidean quantum gravity. *Phys. Rev. D* **54** (1996) 5002, https://doi.org/10.1103/PhysRevD.54.5002
6. Agop M., Buzea C. Gh., Nica P. Local gravitoelectromagnetic effects on a superconductor. *Physica C: Supercond* **339** (2000) 120–128, https://doi.org/10.1016/S0921-4534(00)00340-3
7. Tajmar M., Plesescu F., Seifert B. Measuring the dependence of weight on temperature in the low-temperature regime using a magnetic suspension balance. *Meas. Sci. Technol.* **21** (2010) 015111 (7pp), https://doi.org/10.1088/0957-0233/21/1/015111

8. Tajmar M. Evaluation of enhanced frame-dragging in the vicinity of a rotating niobium superconductor, liquid helium and a helium superfluid. *Supercond. Sci. Technol.* **24** (2011) 125011, https://doi.org/10.1088/0953-2048/24/12/125011
9. Fetisov A.V. Possibility of existing brand-new type of attractive field in $YBa_2Cu_3O_{6+\delta}$. *Phys. C.: Supercond.* **562** (2019) 7–12, https://doi.org/10.1016/j.physc.2019.03.016
10. Fetisov A.V. Distance to massive metal body – a paradoxical parameter that regulates the intensity of the hydration of $YBa_2Cu_3O_{6.75}$. *J. Supercond. Nov. Magn.* **33** (2020) 941–948, https://doi.org/10.1007/s10948-019-05308-0
11. Fetisov A.V. Puzzling behavior of hydrated YBCO. *J. Supercond. Nov. Magn.* **33** (2020) 3341–3348, https://doi.org/10.1007/s10948-020-05620-0
12. Fetisov A.V. Weight balance violation during hydration of $YBa_2Cu_3O_{6+\delta}$. *J. Supercond. Nov. Magn.* **34** (2021) 2725–2732, https://doi.org/10.1007/s10948-021-05979-8
13. Larson A.C. and Von Dreele R.B. GSAS – General Structure Analysis System LANSCE MS-H805, Los Alamos National Laboratory, Los Alamos, NM 87545, 1986.
14. Jorgensen J.D., Veal B.W., Paulikas A.P., Nowicki L.J., Crabtree G.W., Claus H., Kwok W.K. Structural properties of oxygen-deficient $YBa_2Cu_3O_{7-\delta}$. *Phys. Rev. B* **41** (1990) 1863–1877, https://doi.org/10.1103/PhysRevB.41.1863
15. Liang R., Bonn D.A., Hardy W.N. Evaluation of $CuO_2$ plane hole doping in $YBa_2Cu_3O_{6+x}$ single crystals. *Phys. Rev. B* **73** (2006) 180505, https://doi.org/10.1103/PhysRevB.73.180505
16. Ono A. A crystallographic study on $Ba_2YCu_3O_{7-y}$. *Jpn. J. Appl. Phys.* **26** (1987) L1223-L1225, https://doi.org/10.1143/JJAP.26.L1223
17. Andersen N.H., von Zimmermann M., Frello T., Käll M., Mønster D., Lindgård P.-A., Madsen J., Niemöller T., Poulsen H.F., Schmidt O., Schneider J.R., Wolf Th., Dosanjh P., Liang R., Hardy W.N. Superstructure formation and the structural phase diagram of $YBa_2Cu_3O_{6+x}$. *Phys. C.: Supercond.* **317-318** (1999) 259-269, https://doi.org/10.1016/S0921-4534(99)00066-0
18. von Zimmermann M., Schneider J.R., Frello T., Andersen N.H., Madsen J., Käll M., Poulsen H.F., Liang R., Dosanjh P., Hardy W.N. Oxygen-ordering superstructures in underdoped $YBa_2Cu_3O_{6+x}$ studied by hard x-ray diffraction. *Phys. Rev. B* **68** (2003) 104515, https://doi.org/10.1103/PhysRevB.68.104515
19. Yakhou F., Henry J.-Y., Burlet P., Plakhty V.P., Vlasov M., Moshkin S. Oxygen ordering and the 60 K plateau in the ortho-II phase of $YBa_2Cu_3O_{6+x}$ ($0.48 \le x \le 0.62$). *Phys. C.: Supercond.* **333** (2000) 146–154, https://doi.org/10.1016/S0921-4534(00)00184-2
20. Liang R., Bonn D.A., Hardy W.N. Preparation and X-ray characterization of highly ordered ortho-II phase $YBa_2Cu_3O_{6.50}$ single crystals. *Phys. C.: Supercond.* **336** (2000) 57–62, https://doi.org/10.1016/S0921-4534(00)00091-5
21. Leridon B., Monod P., Colson D., Forget A., Thermodynamic signature of a phase transition in the pseudogap phase of $YBa_2Cu_3O_x$ high-$T_c$ superconductor. *A Letters*

*Journal Exploring the Frontiers of Physics* **87** (2009) 17011, https://doi.org/10.1209/0295-5075/87/17011

22. Fetisov A.V., Kozhina G.A., Estemirova S.Kh., Mitrofanov V.Ya. On the room-temperature aging effects in $YBa_2Cu_3O_{6+\delta}$. *Phys. C.: Supercond.* 515 (2015) 54–61, https://doi.org/10.1016/j.physc.2015.05.008
23. Tallon J.L., Bernhard C., Shaked H., Hitterman R.L., Jorgensen J.D. Generic superconducting phase behavior in high-$T_c$ cuprates: $T_c$ variation with hole concentration in $YBa_2Cu_3O_{7-\delta}$. *Phys. Rev. B* **51** (1995) 12911–12914, https://doi.org/10.1103/PhysRevB.51.12911
24. Tallon J.L., Williams G.V.M., Flower N.E., Bernhard C. Phase separation, pseudogap and impurity scattering in the HTS cuprates. *Phys. C.: Supercond.* **282–287** (1997) 236–239, https://doi.org/10.1016/S0921-4534(97)00224-4
25. Veal B.W., Paulikas A.P. Dependence of hole concentration on oxygen vacancy order in $YBa_2Cu_3O_{7-\delta}$: A chemical valence model. *Phys. C.: Supercond.* **184** (1991) 321–331, https://doi.org/10.1016/0921-4534(91)90398-I
26. McCormack R., de Fontaine D. Oxygen configurations and their effect on charge transfer in off-stoichiometric $YBa_2Cu_3O_z$. *Phys. Rev. B* **45** (1992) 12976–12987, https://doi.org/10.1103/PhysRevB.45.12976
27. Poulsen H.F., Andersen N.H., Andersen J.V., Bohr H., Mouritsen O.G. Relation between superconducting transition temperature and oxygen ordering in $YBa_2Cu_3O_{6+x}$. *Letters to Nature* **349** (1991) 594–596, https://doi.org/10.1038/349594a0
28. Matic V.M., Lazarov N.Dj. The origin of the 60 K plateau in $YBa_2Cu_3O_{6+x}$. *J. Phys.: Condens. Matter* **19** (2007) 346230 (9pp), https://doi.org/10.1088/0953-8984/19/34/346230
29. Segawa K., Ando Y. Transport properties of untwinned $YBa_2Cu_3O_y$ single crystals and the origin of the 60-K plateau. *J. Low Temp. Phys.* **131** (2003) 821–830, https://doi.org/10.1023/A:1023427026600
30. Honma T., Hor P.H. Intrinsic electronic superconducting phases at 60 K and 90 K in double-layer $YBa_2Cu_3O_{6+\delta}$. *Phys. Rev. B* **75** (2007) 012508, https://doi.org/10.1103/PhysRevB.75.012508
31. Zaleski T.A., Kopeć T.K. Possible origin of 60-K plateau in the $YBa_2Cu_3O_{6+y}$ phase diagram. *Phys. Rev. B* **74** (2006) 014504 (9pp), https://doi.org/10.1103/PhysRevB.74.014504